\documentclass[10pt,twocolumn,superscriptaddress,aps,pra,showpacs,preprintnumbers,amsmath,amssymb,floatfix,nofootinbib]{revtex4}

\usepackage[latin9]{inputenc}
\usepackage{color}

\usepackage{braket}
\usepackage{graphicx}
\usepackage{esint}
\usepackage[unicode=true,  bookmarks=false,  breaklinks=false,pdfborder={0 0 1},backref=false,colorlinks=true]  {hyperref}

\makeatletter
\@ifundefined{textcolor}{}
{%
 \definecolor{BLACK}{gray}{0}
 \definecolor{WHITE}{gray}{1}
 \definecolor{RED}{rgb}{1,0,0}
 \definecolor{GREEN}{rgb}{0,1,0}
 \definecolor{BLUE}{rgb}{0,0,1}
 \definecolor{CYAN}{cmyk}{1,0,0,0}
 \definecolor{MAGENTA}{cmyk}{0,1,0,0}
 \definecolor{YELLOW}{cmyk}{0,0,1,0}
}

\usepackage[english]{babel}
\usepackage{amsmath}
\usepackage{amssymb}
\usepackage{bm}

\newcommand{\bx}{{\mathbf x}}

\newcommand{\re}{\mbox{Re}}
\newcommand{\im}{\mbox{Im}}

\begin{document}
\title{Fragile fate of driven-dissipative XY phase in two dimensions}
\author{Mohammad F. Maghrebi}
\affiliation{Department of Physics and Astronomy, Michigan State University, East Lansing, Michigan 48824, USA}

\begin{abstract}
Driven-dissipative systems define a broad class of non-equilibrium systems where an external drive (e.g. laser) competes with a dissipative environment. 
The steady state of dynamics is generically distinct from a thermal state characteristic of equilibrium. As a representative example, a driven-dissipative system with a continuous symmetry is generically disordered in two dimensions in contrast with the well-known algebraic order in equilibrium XY phases. In this paper, we study a 2D driven-dissipative model of weakly interacting bosons with a continuous $U(1)$ symmetry. Our aim is two-fold: First, we show that an effectively equilibrium XY phase emerges despite the driven nature of the model, and that it is protected by a natural ${\mathbb Z}_2$ symmetry of the dynamics. Second, we argue that this phase is unstable against symmetry-breaking perturbations as well as static disorder, whose mechanism in most cases has no analog in equilibrium. In the language of renormalization group theory, we find that, outside equilibrium, there are more relevant directions away from the XY phase.

 \end{abstract}

\maketitle

\section{Introduction}

A time-dependent drive continuously pumps energy into a driven system, and eventually heats it up to infinite temperature.
On the other hand, a driven system coupled to a dissipative bath approaches a nontrivial non-equilibrium steady state due the competition between dissipation and external drive. In many-body driven-dissipative systems, the steady state of dynamics may exhibit new, and inherently nonequilibrium, phases. The latter, however, pose a fundamental challenge to our understanding of phases of matter. 

Non-equilibrium systems are, almost by definition, less constrained than their equilibrium counterparts. This implies that, away from equilibrium, dynamics and fluctuations can explore a larger ``phase space''. It is then natural to expect non-equilibrium phases that are not accessible in equilibrium. The converse of this statement could also be true in the sense that a generic equilibrium phase may be non-generic far from equilibrium.
A representative example is a driven-dissipative model with $U(1)$ symmetry in low dimensions.
This model is particularly relevant to driven-dissipative condensates consisting of exciton polaritons in semiconductor quantum wells \cite{Kasprzak06,Lagoudakis08,Roumpos12,Moskalenko2000,Ciuti13-2,Keeling2017superfluidity}.
It has been argued that such driven-dissipative Bose systems in two dimensions cannot exhibit algebraic order, characteristic of the equilibrium XY model, unless they are strongly anisotropic \cite{Altman15}.
This is partly due to the emergence of the Kardar-Parisi-Zhang (KPZ) equation that describes a broad range of driven classical phenomena \cite{Kardar86}.

\begin{figure}[h]
   \centering
    \includegraphics[width=7cm]{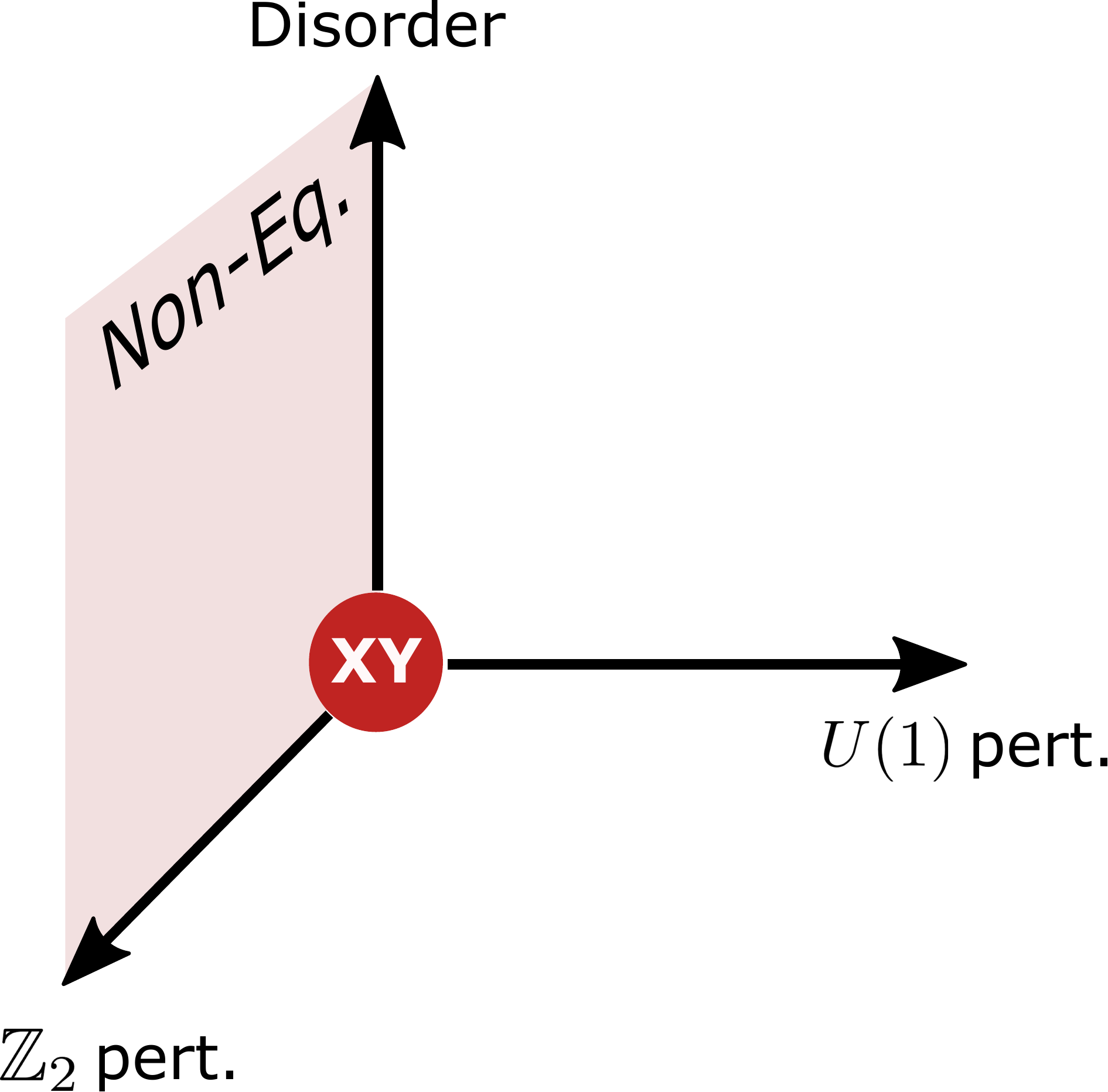}
    \caption{Emergence of an effectively thermal XY phase as the steady state of a non-equilibrium driven-dissipative model with $\mathbb Z_2 \times U(1)$ symmetry of the unit cell. The schematic plot shows relevant perturbations away from the XY phase  which include symmetry-breaking perturbations as well as static disorder. While $U(1)$ symmetry breaking finds an effective equilibrium character, the corresponding mechanisms for $\mathbb Z_2$ symmetry breaking and disorder are of a genuinely non-equilibrium nature. (The highlighted plane represents the subspace spanned by genuinely non-equilibrium perturbations.)}%
    \label{Fig. RG}%
\end{figure}

This manuscript makes a case for the emergence of the XY phase in driven-dissipative systems on the basis of symmetry. We present a case study of a 2D driven-dissipative bosonic model with $U(1)$ symmetry which nevertheless gives rise to an XY phase. We further argue that this is due to the underlying symmetries of the model including an additional $\mathbb Z_2$ symmetry. 
Despite the emergence of the XY phase, the model is shown to be generically unstable to symmetry-breaking perturbations as well as static disorder. We shall argue that, while $U(1)$-symmetry breaking perturbations find a description similar to those in equilibrium, perturbations of the $\mathbb Z_2$ symmetry as well as static disorder are of a genuinely non-equilibrium nature (see Fig.~\ref{Fig. RG}).

The structure of this paper is as follows. In Sec.~\ref{Sec. Model}, we introduce the 2D driven-dissipative model of weakly interacting bosons, and argue on the basis of the Keldysh functional integral that an effectively classical equilibrium XY phase emerges. In Sec.~\ref{Sec. Role of Symmetry}, we undertake a detailed study of the symmetries of the model and the way they constrain the emergent thermodynamic phase. We further discuss perturbations away from symmetries as well as static disorder. Finally, in Sec.~\ref{Sec. Outlook}, we summarize our results and discuss future directions.

\section{Model}\label{Sec. Model}
We consider a driven-dissipative model of weakly interacting bosons on a square lattice in two dimensions. This model is inspired by the spin model introduced in Ref.~\cite{Lee13} and its subsequent treatment in Ref.~\cite{Maghrebi16Keldysh} where spins were mapped to bosons. To define the model, we start from the quantum master equation
\begin{equation}\label{Eq. Master equation}
    \partial_t \rho =-i [\hat H, \rho] +\sum_j \left(\hat L_j \rho \hat L_j^\dagger - \frac{1}{2} \hat L_j^\dagger \hat L_j \rho- \frac{1}{2} \rho \hat L_j^\dagger \hat L_j \right).
\end{equation}
The first term on the right-hand side gives the usual coherent evolution via the Hamiltonian $\hat H$.
The dissipation is subsumed in the second term characterized by the Lindblad operators $L_j$s that describe the incoherent processes. We take the Hamiltonian as
\begin{equation}\label{Eq. Hamiltonian}
   \hat H= J\sum_{\langle i j\rangle} \left(\hat a_i \hat a_j+\hat a_i^\dagger \hat a_j^\dagger\right) + U\sum_j \hat a_j^\dagger \hat a_j^\dagger \hat a_j \hat a_j.
\end{equation}
The first term in the Hamiltonian describes {\it anomalous} hopping between nearest neighbors, while the second term is the on-site interaction.\footnote{For notational convenience, $J$ is defined two times that of Refs.~\cite{Lee13,Maghrebi16Keldysh}.} We can also consider a ``chemical-potential'' term $\sim\sum_j a_j^\dagger a_j$ in the Hamiltonian; the latter, however, does not alter our main conclusions, and will be discussed from the point of view of symmetry in Sec.~\ref{Sec. Role of Symmetry}. Furthermore, we consider weakly interacting bosons where the the interaction ($U$) can be treated perturbatively.
Finally, the incoherent dynamics is given by a single-particle loss
\begin{equation}
  \hat L_j= \sqrt{\Gamma}\, \hat a_j\,.
\end{equation}
The Hamiltonian and the Lindblad operators should be understood in a rotating frame---determined by the frequency of the external drive---after making the rotating wave approximation. The latter is an excellent approximation provided that the drive frequency is much larger compared to other energy scales. We shall not provide a microscopic time-dependent model\footnote{The underlying time-dependent model is not unique, and its explicit form is constrained by experimental feasibility rather than physical principles.}; however, we argue that the driven nature of the dynamics is inherent in the quantum master equation. To this end, note that there is a competition between the Hamiltonian and dissipative dynamics. While the dissipation via $L_j$s favors a state with no particles, or a {\it vacuum}, the Hamiltonian produces pairs of particles out of the vacuum state. The competition between the two gives rise to a steady state at long times with a finite density of particles. This feature has no analog in equilibrium, and is the defining character of driven-dissipative models.

An important property of the model introduced here is that it possesses a $U(1)$ symmetry. To see this, let us consider the checkerboard sublattices $A$ and $B$ of the square lattice. 
The quantum master equation is invariant under the following staggered $U(1)$ transformation
\begin{equation}\label{Eq. sublattice symmetry}
  \hat a_{j\in A} \to e^{i\theta} \hat a_{j\in A},\qquad \hat a_{j\in B} \to e^{-i\theta} \hat a_{j \in B},
\end{equation}
where bosons on the two sublattices are ``rotated'' in opposite directions. This is to ensure that the anomalous hopping in the Hamiltonian remains invariant; all the other terms in the master equation (including the Lindblad terms) are acting on a single site, and respect the symmetry as well. It is then natural to ask whether the continuous $U(1)$ symmetry is broken in the steady state. A mean-field analysis would be a first step to this end (for the spin analog of this model, see Ref.~\cite{Lee13}). However, mean-field-type treatments are at best incomplete since they ignore fluctuations that are crucial to finding the fate of ordered phases in low dimensions. Furthermore, in a nonequilibrium setting, there is even a larger phase space available to dynamics and fluctuations. Instead, we shall follow a field-theory treatment
based on the Keldysh formalism. 

\subsection{Overview of Keldysh formalism \\ and previous results}
The Keldysh formalism adapts the functional-integral techniques to density matrices where two time contours/branches represent the evolution of the bra and ket states in the density matrix. 
In transitioning to the functional integral, the operator $\hat a_j$ is mapped to the fields $a_{j,\pm}(t)$ with the subscripts $\pm$ representing the two branches. The Keldysh functional integral gives a weighted sum (integral) over all configurations of $a_{j,\pm}(t)$.
The weight associated with each configuration is given by the Keldysh action ${\cal S}_K[a_{j,\pm}(t)]$, the form of which is directly determined from the quantum master equation (\ref{Eq. Master equation}). In a coherent-state representation, the Keldysh action can be cast as \cite{Diehl16review}
\begin{equation}
   {\cal S}_K=\int_{t} \left[\sum_j \left(a_{j,+}^* i\partial_t a_{j,+} -a_{j,-}^* i\partial_t a_{j,-}\right) -i {\cal L}\right].
\end{equation}
$\cal L$ contains information about dynamics, and is given by
\begin{align}
  {\cal L}=&-i(H_+-H_-)\\
          \!\!\!+&\sum_j\left[ L_{j,+}L_{j,-}^* -\frac{1}{2} \left( L_{j,+}^*L_{j,+}+L_{j,-}^*L_{j,-}\right)\right] \nonumber
\end{align}
where $H_\pm$ as well as $L_\pm$ contain fields on the $\pm$ contour only.
Clearly, the first line of this equation captures the unitary dynamics, while the second line describes the dissipative dynamics. The particular form of various terms are determined by the simple rule that a term of the form $\hat O \rho \hat O'$ in the quantum master equation translates to $O_+ O'_-$ in the action \cite{DallaTorre13,Sieberer14}.

It is often more convenient to work in the Keldysh basis defined as \cite{KamenevBook,Diehl16review}
\begin{equation}
a_{j,cl}=\frac{a_{j,+}+ a_{j,-}}{\sqrt{2}},\qquad a_{j,q}=\frac{a_{j,+}- a_{j,-}}{\sqrt{2}}.
\end{equation}
This basis is more convenient in separating out the mean value (represented by $a_{cl}$) from the fluctuations around it (due to both $a_{cl}$ and $a_q$ which may be nevertheless of different nature).
Next, we provide a summary of the previous results obtained in the context of a spin model \cite{Lee13,Maghrebi16Keldysh} to the extent that it is relevant to our discussion. Along the way, we also give an overview of the by-now standard techniques and methods.
Motivated by the staggered $U(1)$ symmetry (\ref{Eq. sublattice symmetry}), we allow the order parameter to be different on the two sublattices, but assume that it is uniform within each sublattice. With this assumption, one can take the continuum limit of the lattice model. Following Ref.~\cite{Maghrebi16Keldysh}, we define the bosonic operators on the two sublattices $A$ and $B$ in the continuum as
\begin{equation}
  \hat a_{j\in A} \longrightarrow \hat a(\bx), \qquad \hat a_{j\in B} \longrightarrow \hat b(\bx).
\end{equation}
(With a slight abuse of notation, we have now used $\hat a(\bx)$ to denote the bosonic operators corresponding to the sublattice $A$.)
The corresponding quantum and classical fields associated with the operators $\hat a(\bx)$ and $\hat b(\bx)$ should be identified as $a_{cl/q}(t,\bx)$ and $b_{cl/q}(t,\bx)$. Subsequently, the Keldysh action can be written as a functional of these fields.
It was pointed out in Ref.~\cite{Maghrebi16Keldysh} that one can make the transformation
\begin{align}\label{Eq. New basis}
     \psi_{cl/q}(t, \bx)&=\mp \left[e^{\pm i\pi/4} b_{cl/q}(t, \bx)+e^{\mp i\pi/4} a_{cl/q}^*(t,\bx)\right], \nonumber \\
     \chi_{cl/q}(t,\bx)&=e^{\mp i\pi/4} b_{cl/q}(t,\bx)+e^{\pm i\pi/4} a_{cl/q}^*(t,\bx),
\end{align}
to bring the Keldysh action into a more transparent form at or near the critical point to be further discussed below. This transformation casts the quadratic part of the Keldysh Lagrangian density (the integrand of the space-time integral in the action) as
\begin{align}\label{Eq. Keldsyh 2 Lag}
     {\cal L}_K^{(2)}= \frac{1}{2}\, \Big\{\, &\psi_q^*\big[-\partial_t+J\nabla^2 -r\big] \psi_{cl} +{\rm c.c.} +i\Gamma |\psi_q|^2 \nonumber \\
     +&\chi_q^*\big[-\partial_t -R\big] \chi_{cl} +{\rm c.c.} +i\Gamma |\chi_q|^2 \Big\},
\end{align}
with the constants
\begin{equation}
  r=\Gamma/2-4J, \qquad R= \Gamma/2+4J.
\end{equation}
Importantly, the constant $r$ can be tuned to zero, or {\it criticality}, while $R$ is always finite. Indeed we have used this fact to drop the gradient term acting on $\chi_{cl}$ at long wavelengths. It should be then clear that the critical behavior is captured by $\psi_{cl/q}$, while $\chi_{cl/q}$ are non-critical. At the quadratic level, the two fields are decoupled, and $\chi_{cl/q}$ can be simply dropped; however, interaction mixes the critical and non-critical fields together. We shall not reproduce the interaction terms in the new basis, and refer the interested reader to Ref.~\cite{Maghrebi16Keldysh}. Integrating out $\chi_{cl/q}$ produces an effective interaction term of the form
\begin{equation}\label{Eq. int Lagrangian}
     {\cal L}_K^{\rm int}=-\frac{u}{2}\left(|\psi_{cl}|^2 \psi_{cl}\psi_q^*+{\rm c.c.}\right).
\end{equation}
The (real) coefficient $u \sim U^2 /J$  is obtained via a second-order perturbation theory in the vicinity of the critical point $r=0$ or $J= \Gamma/8$. Of course, a perturbative treatment is justified in the limit of weak coupling $U\ll J$. We remark that there are various nonlinear terms generated in the second-order perturbation theory; however, a simple scaling analysis renders nonlinear terms with higher powers of the quantum field $\psi_q$ irrelevant in the sense of renormalization group (RG) theory. A first step of perturbative RG is to determine scaling dimensions of the fields at the Gaussian fixed point corresponding to the quadratic part of the action.  Demanding that the latter should be scale-invariant at the critical point  under the transformation $\bx\to b \bx $ and $t \to b^2 t$ (relative scaling of space and time coordinates follows from the diffusive nature of the dynamics) requires $\psi_{cl}\to b^0 \psi$ and $\psi_q\to b^2 \psi_q$ in two dimensions. The corresponding scaling dimensions are then $[\psi_{cl}]=0$ and $[\psi_q]=2$. The difference in the scaling dimensions is a consequence of the fact that $\psi_q$ is ``gapped'' in the sense that the action contains the term $\Gamma |\psi_q|^2$ whose coefficient, unlike that of $r\psi_q^* \psi_{cl}+{\rm c.c.}$, cannot be tuned to zero. The relevance of nonlinear terms at the Gaussian fixed point is determined by their RG flow; generally, terms containing higher powers of fields of a larger scaling dimension will be less relevant.

Putting together the quadratic terms in the first line of Eq.~(\ref{Eq. Keldsyh 2 Lag}) with the interaction term in Eq.~(\ref{Eq. int Lagrangian}), we find the effective Keldysh Lagrangian density obtained by integrating out the non-critical fields. Incidentally, the latter can be written as
\begin{equation}\label{Eq. eff Lagrangian}
{\cal L}_K^{\rm eff}=\
\frac{1}{2}\Big\{\psi_q^*\left[-\partial_t \psi_{cl}-{\delta {\cal H}^{\rm eff}}/{\delta \psi_{cl}^*}\right]+{\rm c.c.} +i T^{\rm eff} |\psi_q|^2\Big\},
\end{equation}
where $T^{\rm eff}=\Gamma$, and the functional ${\cal H}^{\rm eff}\equiv {\cal H}^{\rm eff}[\psi_{cl}]$ is given by
\begin{equation}\label{Eq. eff Hamiltonian}
  {\cal H}^{\rm eff}[\psi]=\int_\bx J\,\,|\nabla \psi|^2+r\, |\psi|^2 +u |\psi|^4.
\end{equation}
(A function(al) of $\psi$ should be always interpreted as a function(al) of both $\psi$ and $\psi^*$.) 
The latter has the same form as the Landau-Ginzburg free energy for a complex-valued field $\psi$. The description in terms of the effective action and Hamiltonian can be equivalently cast as a stochastic equation,
\begin{equation}\label{Eq. Langevin eq}
  \partial_t \psi=-\frac{\delta {\cal H}^{\rm eff}}{\delta \psi^*}+\xi(t,\bx),
\end{equation}
where $\xi$ represents a stochastic noise that is correlated as $\langle\xi(t,\bx)\xi^*(t',\bx')\rangle= 2T^{\rm eff} \delta(t-t')\delta(\bx-\bx')$.
Using standard techniques \cite{KardarBook}, one can show that the asymptotic steady state of the effective dynamics in Eq.~(\ref{Eq. Langevin eq}) is given by the probability distribution ($\psi_{cl} \to \psi$)
\begin{equation}
  {\cal P}[\psi]\sim \exp\left(-\frac{{\cal H}^{\rm eff}[\psi]}{T^{\rm eff}}\right),
\end{equation}
which is nothing but a thermal distribution function. 
Despite the nonequilibrium dynamics at the microscopic scale, at long wavelengths, the model effectively behaves as if it is in equilibrium. Of course, the effective Hamiltonian (free energy) and temperature are not in any direct way related to those at the microscopic level.

\subsection{Emergence of XY phase}
An effective classical and equilibrium behavior opens up the sophisticated toolbox of statistical mechanics. In the context of the model considered here, one can immediately draw intuition from the classical XY model in two dimensions. In particular, {\it vortices} should be properly taken into account. To this end, let us take $\psi_0=\langle\psi_{cl}\rangle$, and define $K\equiv 2J|\psi_0|^2/\Gamma$ which is commonly known as spin stiffness. It is a classical result due to Kosterlitz and Thouless that a quasi-long-range order with algebraic decay of correlations emerges when \cite{KardarBook}
\begin{equation}\label{Eq. constraint}
  K>\frac{2}{\pi}.
\end{equation}
In the opposite regime where this constraint is not satisfied, vortices proliferate destroying the algebraic order and leading to an exponential decay of correlations.

For the spin model introduced in Ref.~\cite{Lee13}, it was shown that the constraint (\ref{Eq. constraint}) cannot be satisfied \cite{Maghrebi16Keldysh}, and the XY phase will not be realized.This is because $\psi_0$ represents the expectation value of a spin operator which may be saturated, and consequently $K\sim|\psi_0|^2$ would not be large enough. In contrast, in our bosonic model, $|\psi_0|$ can be arbitrarily large in favor of the constraint (\ref{Eq. constraint}). The saddle-point approximation of Eq.~(\ref{Eq. eff Hamiltonian}) yields
\begin{equation}
  |\psi_0|^2\approx \frac{|r|}{2u},
\end{equation}
when $r<0$. Recalling that $u\sim U^2/J$, the constraint (\ref{Eq. constraint}) is easily satisfied in the weak-coupling regime $U\ll J$. 
We thus conclude that an XY phase is realized in the driven-dissipative model of weakly interacting bosons introduced here.

Our perturbative treatment still lacks an important discussion. We have used second-order perturbation theory to show that the resulting action takes a form that finds a description in terms of an effective free energy. It is important, however, to show that this is not an artifact of our approximation, but rather is protected by the symmetries of the model. This is particularly important in the context of a 2D driven-dissipative model with $U(1)$ symmetry where the XY phase is shown to be generically unstable to KPZ-like physics \cite{Altman15}.
In the next section, we discuss the symmetries of the model, and argue that the XY phase is indeed protected by these symmetries. Furthermore, we show that relaxing these symmetries generically tends to destroy the XY phase.

\section{Role of Symmetry}\label{Sec. Role of Symmetry}
In  addition to the $U(1)$ symmetry (Eq.~(\ref{Eq. sublattice symmetry})), the model defined in the previous section has a ${\mathbb Z}_2$ symmetry under sublattice exchange $A \leftrightarrow B$. 
Hence, in a unit cell consisting of two sites (one from each sublattice), we have an enlarged $\mathbb Z_2\times U(1)$ symmetry. In the continuum, the $\mathbb Z_2$ symmetry interchanges the fields,
\(
  \hat a(\bx) \leftrightarrow \hat b(\bx),
\)
which constitutes a fundamental symmetry of the model beyond any approximation or perturbation scheme. The latter directly translates to a transformation in terms of the fields, $a_{cl/q}(t,\bx) \leftrightarrow b_{cl/q}(t,\bx)$, as a symmetry of the Keldysh action ${\cal S}_K[a_{cl/q},b_{cl/q}]$. In the basis of the fields $\psi_{cl/q}$ and $\chi_{cl/q}$ defined in Eq.~(\ref{Eq. New basis}), this symmetry simply reads as complex conjugation,
\begin{equation}
\begin{split}
  \psi_{cl/q}(t,\bx) &\leftrightarrow \psi^*_{cl/q}(t,\bx), \\
  \chi_{cl/q}(t,\bx) &\leftrightarrow \chi^*_{cl/q}(t,\bx).
\end{split}
\end{equation}
Being a fundamental symmetry of the model, this transformation should also be a symmetry of the effective Keldysh action ${\cal S}^{\rm eff}_K[\psi_{cl/q}]$ obtained by integrating out the non-critical fields $\chi_{cl/q}$,
\begin{equation}\label{Eq. symmetry constraint}
  {\cal S}^{\rm eff}_K[\psi_{cl}^*,\psi_{q}^*]={\cal S}^{\rm eff}_K[\psi_{cl},\psi_{q}].
\end{equation}
This equation imposes a strong constraint on the form of the Keldysh action. To fully exploit it, we also note that a general Keldysh action ${\cal S}_K[a_{cl}, a_{q}]$ (with $a_{cl/q}$ representing {\it all} the fields with quantum numbers suppressed) always satisfies
\(
  {\cal S}^*_K[a_{cl}, a_{q}]=-{\cal S}_K[a_{cl}, -a_{q}].
\)
This equation follows from the causal structure of the Keldysh action, and ensures that $(G^{R})^\dagger=G^A$ and $(G^K)^\dagger=-G^K$ where $G^{R,A,K}$ are retarded, advanced, and Keldysh Green's functions, respectively. It then follows that $\re({\cal S}_K[a_{cl}, a_{q}])$ is odd in $a_q$ while $\im({\cal S}_K[a_{cl}, a_{q}])$ is even in $a_q$. For future reference, we specialize the causality structure to the effective Keldysh action,
\begin{equation}\label{Eq. General property specialized}
  \left({\cal S}^{\rm eff}_K[\psi_{cl}, \psi_{q}]\right)^*=-{\cal S}^{\rm eff}_K[\psi_{cl}, -\psi_{q}].
\end{equation}

Using the $\mathbb Z_2\times U(1)$ symmetry, we next expand the Keldysh action in classical and quantum fields, and only keep spatial and time derivatives to the lowest order. With the exception of the classical field, this is well justified due to the corresponding scaling dimensions ($[\partial_t]=2$, $[\partial_\bx]=1$, and $[\psi_q]$=2). However, a similar scaling argument fails for the classical field since $[\psi_{cl}]=0$ at the critical point. Moreover, we need to consider a parameter regime at a finite distance away from the critical point where Eq.~(\ref{Eq. constraint}) is satisfied and the classical field assumes a finite value. We will present a more general argument below; for now, we simply expand the action in both classical and quantum fields as\footnote{The expansion starts at the linear order in quantum field since ${\cal S}_{K}[a_{cl},a_q=0]=0$ as a general property of the Keldysh action \cite{KamenevBook}.}
\begin{align}\label{Eq. S-K expanded}
  {\cal S}^{\rm eff}_K=\int_{t,\bx} &\psi_q^*\left[-Z\partial_t\psi_{cl} -\tilde J\nabla^2 \psi_{cl}-\partial V/\partial{\psi_{cl}^*}\right]+{\rm c.c.} \nonumber \\
  &+i \tilde \Gamma |\psi_q|^2+\cdots.
\end{align}
$V(\psi_{cl})$ is a function of the modulus $|\psi_{cl}|$, and can be expanded as $V(\psi)=\tilde r|\psi|^2+\tilde u |\psi|^4+\cdots$. Note that the action is at most quadratic in the quantum field, but possibly contains higher-order terms in the classical field. Also it is written in a way that explicitly respects the $U(1)$ symmetry. Furthermore,
the complex conjugation in the first line of the action is to ensure the reality of $\psi_q$-odd terms that follows from Eq.~(\ref{Eq. General property specialized}). In general, the coefficients $Z, \tilde J, \tilde r, \tilde u,\cdots$ can be complex-valued ($\tilde \Gamma$ has to be real on the basis of Eq.~(\ref{Eq. General property specialized})). However, the symmetry constraint in Eq.~(\ref{Eq. symmetry constraint}) ensures that all the coefficients are real. Therefore, the action can be written in a form consistent with Eqs.~(\ref{Eq. eff Lagrangian}, \ref{Eq. eff Hamiltonian}) with a (real-valued) Hamiltonian, which is what we wanted to show.

It is instructive to present a more general argument within the XY phase. Let us first represent the classical and quantum fields as
\begin{equation}\label{Eq. density-phase rep}
  \psi_{cl}=\sqrt{\rho_0+\pi}\, e^{i\theta}, \quad \psi_q=\zeta e^{i\theta},
\end{equation}
where $\rho_0\equiv |\psi_0|^2=\langle|\psi_{cl}|^2\rangle $ is the average density in the ordered phase, $\pi$ characterizes density fluctuations, and $\zeta=\zeta_1+i\zeta_2$ is a complex field representing the quantum field. Here we have followed the notation in Ref.~\cite{Diehl16review} in factoring out a common phase factor from both classical and quantum fields. The symmetry constraint (\ref{Eq. symmetry constraint}) in the new basis reads
\begin{equation}\label{Eq. symmetry constraint new basis}
  {\cal S}_K^{\rm eff}[\pi, \theta,\zeta_1,\zeta_2]={\cal S}_K^{\rm eff}[\pi, -\theta,\zeta_1,-\zeta_2].
\end{equation}
Moreover, the $U(1)$ symmetry requires the action to be invariant under $\theta \to \theta +{\rm const}$.
With these constraints on the form of the Keldysh action together with Eq.~(\ref{Eq. General property specialized}), we can write the most general Keldysh action as\footnote{Again, the expansion starts at the linear order in $\zeta$ representing the quantum field. A linear term in $\zeta_1$ is also allowed, but nevertheless can be absorbed in a redefinition of $\pi$, or equivalently a renormalization of the density $\rho_0$.}
\begin{align}\label{Eq. S-K density phase}
  {\cal S}_K^{\rm eff}=\int_{t,\bx} &\zeta_2\left(- Z'\partial_t \theta-J'\nabla^2 \theta\right)+J''\zeta_1 (\nabla\theta)^2 -u'\zeta_1 \pi \nonumber \\
  &+i \Gamma'\zeta_1^2+i\Gamma'' \zeta_2^2+\cdots,
\end{align}
where the ellipses represent irrelevant terms that contain higher powers of spatial and time derivatives or the fields $\pi$, $\zeta_1$ and $\zeta_2$ due to the corresponding scaling dimensions ($[\pi]=[\zeta_1]=[\zeta_2]=2$). All the coefficients ($Z',K',K'',u',\Gamma', \Gamma''$) in the action are real as a consequence of the causality structure in Eq.~(\ref{Eq. General property specialized}). We note that, unlike Eq.~(\ref{Eq. S-K expanded}), we have not made an expansion in powers of $\psi_{cl}$ which in the ordered phase can be possibly large. For the special case of the action in Eq.~(\ref{Eq. S-K expanded}), we have $Z'\sim \sqrt{\rho_0} Z$, $K'\sim K''\sim \sqrt{\rho_0}\tilde J$, $u'\sim \sqrt{\rho_0}u$, and $\Gamma'=\Gamma''=\tilde\Gamma$. Now note that the integration over $\pi$ in Eq.~(\ref{Eq. S-K density phase}) gives a delta function that sets $\zeta_1=0$, and the Keldysh action can be simply written as
\begin{equation}
  {\cal S}_K^{\rm eff}=\int_{t,\bx} \zeta_2\left(- Z'\partial_t \theta-K'\nabla^2 \theta\right)+i\Gamma'' \zeta_2^2,
\end{equation}
where irrelevant terms are simply dropped.
It is then straightforward to see that the steady state is given as an effective thermal distribution with the partition function
\begin{equation}
  \int D\theta \,\,\exp\left[-\frac{K}{2}\int(\nabla\theta)^2\right],
\end{equation}
corresponding to the XY Hamiltonian with the spin stiffness $K=K'Z'/\Gamma''$.

Having shown that a description in terms of the XY Hamiltonian is guaranteed by the symmetries of the model, we next turn our attention to symmetry-breaking perturbations. We shall see that the XY phase is generically unstable to such perturbations (see Fig.~\ref{Fig. RG}).

\subsection{Perturbing $U(1)$ symmetry}
There are a number of ways that $U(1)$ symmetry can be explicitly broken. A representative example is nearest-neighbor hopping,
\begin{equation}
  \widehat{\Delta\!H}= \alpha \sum_{\langle i j\rangle}\hat a_i^\dagger \hat a_j+{\rm H.c.}
\end{equation}
Since neighboring sites belong to different sublattices, the $U(1)$ symmetry in Eq.~(\ref{Eq. sublattice symmetry}) is explicitly broken. In the spin analog of Ref.~\cite{Lee13}, this amounts to having $J_x \approx -J_y$ with a slightly different $|J_x|$ and $|J_y|$ of the corresponding XX and YY interactions. In the continuum, we have $\widehat{\Delta\!H}\propto\alpha\int_\bx \hat a^\dagger (\bx) \hat b(\bx)+{\rm H.c.}+\cdots$, and the corresponding term in the action reads $\Delta {\cal S}\propto\alpha\int_{t,\bx} a_q^* b_{cl}+a_{cl}^*b_q+{\rm c.c.}+\cdots$ with the ellipses indicating the less relevant terms. Writing the latter in the basis of $\psi_{cl/q}$ and $\chi_{cl/q}$ and integrating out the non-critical fields $\chi_{cl/q}$, we find, to leading order in $U/J$,
\begin{align}\label{Eq. delta S}
  \Delta {\cal S}_{K}^{\rm eff}\propto \frac{\alpha U}{J}\int_{t,\bx}&\left[\psi_{q}^*\left(\psi_{cl}^*+|\psi_{cl}|^2\psi_{cl}^*
  + \psi_{cl}^3\right)+{\rm c.c.}\right].
\end{align}
All the terms reported here explicitly break the $U(1)$ symmetry, while those that simply renormalize terms already present in the non-perturbed action as well as the less relevant terms are omitted. Also, we have not kept track of relative coefficients (of order 1) between different terms.
The first term under the integral in the effective action ($\psi_q^* \psi_c^*+{\rm c.c.}$) can be cast as a correction to the effective Hamiltonian in Eq.~(\ref{Eq. eff Hamiltonian}) as
\begin{equation}
  \Delta\!{\cal H}^{\rm {eff}}_{1}\propto \frac{\alpha U}{J} \int_\bx \psi^2 +(\psi^*)^2,
\end{equation}
with the subscript denoting the corresponding term. This expression is nothing but a perturbation of the XY model familiar in the context of statistical physics. In the ordered phase, $\psi_{cl} \approx \sqrt{\rho_0} \,e^{i\theta}$, the correction to the effective Hamiltonian becomes 
\begin{equation}\label{Eq. cos 2 theta}
\Delta\!{\cal H}^{\rm {eff}}_{1}\propto \frac{\alpha \rho_0 U}{J} \int_\bx \cos (2\theta).
\end{equation}
Generally, a cosine perturbation of the from $\cos(p\theta)$ is irrelevant if $K< p^2/(8\pi)$ \cite{Nelson77}. However, the latter cannot be satisfied due to the condition (\ref{Eq. constraint}), and thus the cosine perturbation grows under RG pinning the value of $\theta$. The XY phase and its characteristic algebraic order will be then destroyed at long wavelengths.

On the other hand, the second and the third terms under the integral in Eq.~(\ref{Eq. delta S}) cannot be derived from a Hamiltonian.\footnote{A term in the Keldysh action of the form $\psi_q^* f(\psi_{cl}) +{\rm c.c.}$ can be cast as $\psi_q^* \delta {\cal H}/\delta\psi_{cl}^* +{\rm c.c.}$---in a fashion similar to Eq.~(\ref{Eq. eff Lagrangian})---if the function $f(\psi)$ satisfies the condition $\partial f/\partial \psi=\partial f^*/\partial \psi^*$.} 
To treat these terms, we resort to the density-phase representation of Eq.~(\ref{Eq. density-phase rep}). In this representation, the $U(1)$ perturbation leads to a correction to the action of the form (relative and overall coefficients are neglected)
\begin{equation}
  {\Delta {\cal S}}_K^{\rm eff}\propto \int_{t,\bx} \zeta_1\cos(2\theta)+\zeta_2\sin (2\theta).
\end{equation}
This action contains the most relevant perturbations that also respect the ${\mathbb Z}_2$ symmetry (Eq.~(\ref{Eq. symmetry constraint new basis})).
Now the first term under the integral simply drops since the functional integration over $\pi$ sets $\zeta_1=0$. The second term perturbs the $U(1)$ symmetry, but can be similarly cast as a correction to the effective Hamiltonian as in Eq.~(\ref{Eq. cos 2 theta}).

In short, a perturbation of the $U(1)$ symmetry of the form considered in this section can be simply considered as a perturbation of the XY Hamiltonian, which, using standard techniques, can be shown to destroy the XY phase at long wavelengths.

\subsection{Perturbing ${\mathbb Z}_2$ symmetry}
In this section, we study the consequences of breaking the ${\mathbb Z}_2$ sublattice symmetry. Naively, this symmetry can be broken by adding to the Hamiltonian a staggered chemical potential (with different chemical potentials on the two sublattices),
\begin{equation}
  \mu_A\sum_{j\in A} \hat a_{j}^\dagger \hat a_{j}+ \mu_B \sum_{j\in B}\hat a_j^\dagger \hat a_j.
\end{equation}
However, one can remove the asymmetry by exploiting a gauge transformation via the unitary operator
\begin{equation}\label{Eq. unitary}
   \hat{\cal U}(t)=\exp\left[-i\mu t\Big(\sum_{j\in A}\hat a_{j}^\dagger \hat a_{j}-\sum_{j\in B}\hat a_j^\dagger \hat a_j\Big)\right].
\end{equation}
In an appropriate rotating frame with $\mu=(\mu_A-\mu_B)/2$, the above perturbation can be brought into a form with $\mu_A'=\mu_B'$. The latter satisfies all the symmetries ($\mathbb Z_2\times U(1)$ as well as the translation symmetry) of the model, and thus 
only slightly renormalizes the effective Hamiltonian.
It is instructive to view this argument in the basis of the Keldysh action. The corresponding correction to the effective Keldysh action is given by
\begin{equation}\label{Eq. Keldysh chemical potential}
  {\Delta {\cal S}}_K^{\rm eff} = i\frac{\mu_A-\mu_B}{4} \int_{t,\bx} \left(\psi_{q}^* \psi_{cl}-{\rm c.c.}\right).
\end{equation}
Naively, this term breaks the ${\mathbb Z}_2$ symmetry since it is not invariant under $\psi_{cl,q} \leftrightarrow \psi_{cl,q}^*$. However, the stochastic equation that follows from the Keldysh action reads ($\psi_{cl}\to \psi$)
\begin{equation}\label{Eq. Langevin chemical potential}
  \partial_t \psi=-\frac{\delta {\cal H}_{\rm eff}}{\delta\psi^*} +i \frac{\mu_A-\mu_B}{2} \psi+\xi(t,\bx).
\end{equation}
Making the transformation $\psi \to \psi e^{it(\mu_A-\mu_B)/2}$, we recover the unperturbed form of the stochastic equation that can be equivalently described by an effective Hamiltonian, consistent with the gauge transformation in Eq.~(\ref{Eq. unitary}).

To explicitly break the $\mathbb Z_2$ sublattice symmetry, we take the decay rates to be slightly different on the two sublattices as
\begin{equation}
  \hat L_{j\in A}=\sqrt{\Gamma_{A}} \, \hat a_j, \quad \hat L_{j\in B}=\sqrt{\Gamma_{B}} \, \hat a_j,
\end{equation}
with $\Gamma_{A/B}= \Gamma \pm \Delta \Gamma$. We stress that any generic perturbation of the $\mathbb Z_2$ symmetry should also lead to the same conclusions. Writing the corresponding Keldysh action and integrating out the non-critical fields, we find an effective Keldysh action, to the first nontrivial order in $U/J$, as
\begin{align}\label{Eq. delta S 2}
  \Delta {\cal S}_{K}^{\rm eff}\propto \frac{i\Delta\Gamma \,U}{J}\int_{t,\bx}&\left(\psi_{q}^*\psi_{cl}+\psi_{q}^*|\psi_{cl}|^2\psi_{cl} -{\rm c.c.}\right).
\end{align}
(We have not kept track of relative coefficients.) The first term under the integral can be gauged away by going to a rotating frame similar to Eq.~(\ref{Eq. Keldysh chemical potential}); however, the second term cannot be dealt with in a similar fashion. Indeed adding the latter to the non-perturbed Keldysh action, we find a renormalized interaction term $\sim u_{\rm ren} \psi_q |\psi_{cl}|^2 \psi + {\rm c.c.}$ with a complex-valued coefficient $u_{\rm ren}=u'_{\rm ren}+iu''_{\rm ren}$. As shown in Refs.~\cite{Altman15}, this feature generically leads to the KPZ equation, and takes us outside the effective equilibrium description. 
The emergence of the KPZ equation can be generally argued on the basis of symmetry. In the absence of the $\mathbb Z_2$ symmetry, many new terms are allowed in the Keldysh action.
One such term is $\zeta_2 (\nabla\theta)^2$, a term that was previously disallowed due to the symmetry under the simultaneous transformation $\zeta_2\to -\zeta_2$ and $\theta \to -\theta$.\footnote{Other generated terms, $\zeta_1 \nabla^2\theta$, $\zeta_2 \pi$, and $\zeta_1 \zeta_2$, also lead to the same qualitative behavior.}
The inclusion of the latter term in Eq.~(\ref{Eq. eff Lagrangian}) leads to the KPZ equation
\begin{equation}
  \partial_t\theta=J\nabla^2\theta +\lambda(\nabla\theta)^2+\eta(t,\bx),
\end{equation}
where $\eta$ represents a (real-valued) stochastic noise that is correlated as $\langle\eta(t,\bx)\eta(t',\bx')\rangle= (\Gamma/\sqrt{\rho_0}) \delta(t-t')\delta(\bx-\bx')$.
The coefficient of the nonlinear term ($\lambda$) vanishes with the perturbation ($\sim\Delta\Gamma$) away from the $\mathbb Z_2$ symmetry.
Notice that this term cannot be derived from a Hamiltonian in a similar fashion as Eq.~(\ref{Eq. Langevin eq}). At the Gaussian fixed point (ignoring the compact nature of $\theta$), the scaling dimension $[\theta]=0$, and the new term in the KPZ equation is marginal \cite{KardarBook}. To a higher order in perturbation theory, the latter term can be shown to be marginally relevant, and leads to a stretched-exponential decay of the correlation function and the destruction of the XY phase \cite{Altman15}. The interested reader is referred to Refs.~\cite{Diehl16review,Sieberer16cKPZ,Keeling2017superfluidity} for more details on the emergence of the KPZ equation in driven-dissipative condensates.

\subsection{Random disorder}
In this section, we consider the effect of disorder on the behavior of our model. In a disordered system, translation symmetry is broken at a microscopic level, which nevertheless is restored by an ensemble average over disorder configurations. A generic example is a disordered chemical potential
\begin{equation}\label{Eq. Random disorder Hamiltonian}
  \widehat{\Delta\!H}=\sum_j {\mu_j}\hat a_j^\dagger \hat a_j,
\end{equation}
where $\mu_j$ on each site is a static random variable drawn from a Gaussian distribution.
Unlike a staggered chemical potential (see the previous subsection), a disordered chemical potential cannot be gauged away. Carrying out the same steps of writing the perturbation in the continuum, and integrating out noncritical fields, we find a correction to the effective action, to the zeroth order in $U/J$, as
\begin{equation}
    {\Delta \cal {\cal S}}_K^{\rm eff} =\frac{i}{2}\int_{t,\bx} \upsilon \left(\psi_q^* \psi_{cl}-{\rm c.c.}\right),
\end{equation}
where $\upsilon\equiv \upsilon(\bx)$ is correlated as $\langle \upsilon(\bx)\upsilon(\bx')\rangle=\varkappa\delta(\bx-\bx')$ with $\varkappa$ the disorder strength.
Disorder superficially breaks the $\mathbb Z_2$ symmetry (Eq.~(\ref{Eq. symmetry constraint})); however, the integral over the Gaussian distribution restores this symmetry (in the same way that translation symmetry is restored in a disordered system). This can be more precisely formulated as a modified symmetry under the transformation in Eq.~(\ref{Eq. symmetry constraint}) together with $\upsilon(\bx) \to -\upsilon(\bx)$. It is more convenient to cast the above equation in the density-phase representation of Eq.~(\ref{Eq. density-phase rep}) to find
\(
  {\Delta \cal {\cal S}}_K^{\rm eff} =\sqrt{\rho_0}\int_{t,\bx} \zeta_2 \upsilon.
\)
This term satisfies the $\mathbb Z_2$ symmetry which, in this representation, is defined as the symmetry under $\theta \to -\theta$, $\zeta_2\to-\zeta_2$, and $\upsilon\to-\upsilon$ as a close analog of Eq.~(\ref{Eq. symmetry constraint new basis}). Indeed this is the only relevant correction to the action that involves $\upsilon$. This follows from the scaling dimension of static disorder, $[\upsilon]=1$, determined from its Gaussian distribution.\footnote{Note that the disorder average can be performed at the level of the Keldysh functional integral since the partition function is normalized to $Z=1$ by construction. This allows us to directly compare the terms in the action to the disorder distribution \cite{KamenevBook}.} Putting all the relevant terms together in the Keldysh action after integrating out $\pi$ and setting $\zeta_1=0$, one finds 
\begin{equation}
  {\cal {\cal S}}_K^{\rm eff} =\int_{t,\bx} \sqrt{\rho_0}\zeta_2\left(-\partial_t\theta+J\nabla^2\theta+\upsilon\right)+\frac{i \Gamma}{2}\zeta_2^2.
\end{equation}
Importantly the term depending on $\upsilon$ cannot be cast as the functional derivative of a proper potential term (a naive guess $\upsilon \theta$ does not respect the gauge freedom $\theta \to \theta+{\rm const.}$). It is instructive to write the corresponding Langevin equation,
\begin{equation}
  \partial_t \theta=J\nabla^2\theta+\upsilon(\bx)+\eta(t,\bx).
\end{equation}
Note that $\eta$ represents white noise, while $\upsilon$ denotes delta-function-correlated static disorder.
Clearly, the latter cannot be gauged away by going to a rotating frame due to its spatial dependence. Furthermore, static disorder, being perfectly correlated in time, should be expected to dominate over white noise. This is indeed the case, and is easily seen on the basis of scaling analysis. A convenient way to see this is to obtain the disorder-averaged Keldysh action by integrating over $\upsilon$,
\begin{align}
   \overline{{\cal S}_K^{\rm eff}} =&\int_{t,\bx} \left[\sqrt{\rho_0}\zeta_2\left(-\partial_t\theta+J\nabla^2\theta\right) + \frac{i \Gamma}{2}\zeta_2^2\right]\nonumber\\
   +&\frac{i\rho_0\varkappa }{2}\int_{t, t',\bx} \zeta_2(t,\bx)\zeta_2(t',\bx),
\end{align}
where the double time integral in the last line runs from $-\infty$ to $+\infty$ for both $t$ and $t'$.
With $[\zeta_2]=2$ at the XY fixed point described by the first line of this equation, a simple power-counting analysis reveals that $\varkappa$ grows under RG as
\begin{equation}
  \frac{d\varkappa}{dl}=2\varkappa.
\end{equation}
Therefore, static disorder takes the system into a disordered phase at long wavelengths.

We stress that the instability to static disorder discussed here is a purely non-equilibrium phenomenon. Alternatively, imagine that the effect of disorder could be absorbed in a correction to the effective Hamiltonian of the form $\Delta {\cal H}^{\rm eff} \sim\int_\bx \tilde \upsilon|\psi|^2$ with $\tilde \upsilon(\bx)$ a static random potential; in fact, the disorder potential in Eq.~(\ref{Eq. Random disorder Hamiltonian}) produces such a correction at a higher order in $U/J$.  Nevertheless, various terms generated from the disordered effective Hamiltonian can be shown to either vanish or become irrelevant in the sense of RG. Crucially, it is the non-equilibrium nature of the disorder potential that is responsible for the destruction of the XY phase.

\section{Summary and outlook}\label{Sec. Outlook}
In this paper, we have considered a driven-dissipative model of weakly interacting bosons with $U(1)$ symmetry in two dimensions. We have shown that an effectively classical equilibrium XY phase emerges as the steady state despite the driven nature of the model. The emergence of the XY phase has been argued on the basis of an additional $\mathbb Z_2$ symmetry due to the sublattice exchange of the lattice model. Various perturbations of symmetry as well as static disorder have been considered, against which the XY phase is shown to be unstable. It is further argued that $\mathbb Z_2$ symmetry-breaking perturbations as well as static disorder are genuinely of nonequilibrium nature, perturbing the XY phase in directions that are not accessible in equilibrium. More generally, nonequilibrium systems allow for new types of dynamics and fluctuations, which should be properly taken into account in order to determine the nature of phases and phase transitions in the thermodynamic limit. A natural question for future study is the fate of this model in the limit of strong coupling (large $U$). The perturbative arguments presented in this manuscript are not directly applicable in this limit. It is also interesting to consider other types of symmetry ($O(n)$ symmetry, for example), and to compare and contrast the emergent behavior in and out of equilibrium on the basis of symmetry. It would be worthwhile to identify additional symmetries, if any, that constrain the corresponding driven-dissipative models to exhibit an effectively equilibrium behavior.

\section{Acknowledgements}
We acknowledge Sergey V. Syzranov for a critical reading of the manuscript. The author acknowledges start-up funding from Michigan State University.

%


\end{document}